# Ghosts! A Location-Based Bluetooth LE Mobile Game for Museum Exploration

*Tommy Nilsson, Alan F. Blackwell, Carl Hogsden and David Scruton*

**Abstract**
BLE (Bluetooth Low Energy) is a new wireless communication technology that, thanks to reduced power consumption, promises to facilitate communication between computing devices and help us harness their power in environments and contexts previously untouched by information technology. Museums and other facilities housing various cultural content are a particularly interesting area of application. The University of Cambridge Museums consortium has put considerable effort into researching the potential uses of emerging technologies such as BLE to unlock new experiences enriching the way we engage with cultural information. As a part of this research initiative, our ambition has been to examine the challenges and opportunities introduced by the introduction of a BLE-centred system into the museum context. We present an assessment of the potential offered by this technology and of the design approaches that might yield the best results when developing BLE-centred experiences for museum environments. A pivotal part of our project consisted of designing, developing and evaluating a prototype mobile location-based BLE-centred game. A number of technical problems, such as unstable and fluctuating signal strength, were encountered throughout the project lifecycle. Instead of attempting to eliminate such problems, we argued in favour of embracing them and turning them into a cornerstone of the gameplay.
Our study suggested that this alternative *seamful* design approach yields particularly good results when deploying the technology in public environments. The project outcome also demonstrated the potential of BLE-centred solutions to reach out and engage new demographics, especially children, extending their interest in museum visits.

**Key Words:** Bluetooth LE, human computer interaction, game design, internet of things, museum studies, ubiquitous computing, University of Cambridge Museums consortium, user-centered design.

\*\*\*\*\*

1. **What Is Bluetooth LE?**

   Information technologies are currently in the process of moving away from the desktop in order to embrace a more convenient and practical paradigm in the form of mobile computing. Whereas traditional desktop computers restricted our interactions with digital data to taking place within relatively constrained time and space, such as while at work or at home, the fact that we are now increasingly often carrying computing devices with us is turning our immersion with digital information into a permanent state of our lives.

Moreover, a range of technological innovations, such as QR (quick response) codes or NFC (near field communication), taking advantage of and improving the context awareness of our mobile devices, is emerging and subsequently enabling us to interact with our physical surroundings. In parallel with this trend, the decreasing size and cost of the microchip is allowing computational capabilities to be implemented directly into a growing spectrum of our daily life tools in order to augment their functionality, resulting in a plethora of emerging products such as smart shoes or smart fridges.[1] By complementing this computational capability with wireless connectivity, these smart objects can communicate with each other to operate in a coordinated manner. Commentators are increasingly using the term *internet of things*, to describe this network of wirelessly interconnected objects enhanced through computational capabilities. Due to their portability and the context awareness, mobile devices are the most appropriate interface facilitating our interactions with the internet of things.

However, this trend has been hampered by high battery consumption of wireless connectivity and a resulting short usage time. The new Bluetooth Low Energy wireless communication standard (BLE), addresses this issue. Engineered with power efficiency in mind, a BLE beacon can reportedly operate for up to two years on a single coin battery,[2] considerably increasing the operational capabilities of smart objects. BLE can thus play a significant role in advancing the internet of things from a mere futuristic vision to a mainstream level of use.

The potential benefits of this technology have sparked interest in a range of fields, such as marketing and entertainment. For example, by attaching a BLE beacon onto a physical object and programming it to transmit identification information about that particular object, every mobile user moving into its proximity can receive an automated message notifying of the object's presence.

Besides obvious areas of application, such as location-based advertising, this practice might be particularly useful in a museum environment where a similar system could be relatively easily used to provide visitors with additional information about the artefacts in an exhibition. Moreover, the Cambridge Museums consortium has also explored the potential of BLE technology to engage younger audiences with content extending beyond traditional museum experiences. We therefore decided to centre our project on the development of a location-based BLE-centred mobile game that would merge the entertainment value of exploration games with the traditional educational and cultural value offered by museums.

Although focussing on the specific case study of museum environments, a broader concern of our project was to highlight the problems and possible design solutions relevant to the process of integrating a BLE-based mobile system into public environments in general.

_________________________________________________________________

## 2. University of Cambridge Museums

The University of Cambridge Museums consortium consists of the eight museums within the University, working in partnership also with the University's Botanic Garden and other University collections. The museums' world class collections range across art, antiquities, archaeology, anthropology, natural history and the history of science. Major Partner Museum programme funding from the Arts Council has enabled the museums to explore new ways of creating access to its collections to attract new audiences and to make innovative use of digital technology as part of this strategy.

The University Museums are all within walking distance of the city centre and making connections between the collections and guiding visitors from one museum to another has been a strand of the consortium's work. In this context BLE beacons offer the potential to lead visitors through the gallery spaces and across different museums. The University Museums have used non-digital games in the past to link the collections, notably a card-collecting game that featured museum objects.[3] The idea of developing a digital equivalent to that game which could be played on a visitor's smartphone was a starting point for this project.

Triggering content on a mobile device is not new to museums. Various technologies have been used, but usually have depended on an action performed by the visitor such as scanning a QR code or keying in an ID number. The Fitzwilliam Museum first experimented with tagging exhibits using infra-red transmitters developed by a local company *Hypertag* (a spin-out from the University) in 2002.[4] The powered infra-red tags were later integrated into the object labels and the signal picked up on handheld computers hired out to visitors. Although a core feature of museum exhibition practice is to deliver interpretative content whilst visitors are standing in front of specific objects, achieving this via automatic triggering (using triangulation of signals for example) has been less successful. Nevertheless, triggering content at a broader level, such as when entering a distinct gallery space within a museum, can be useful in providing both interpretative and navigational information to the user. There is a role perhaps for BLE beacons in museums for this less granular approach, with beacons deployed to help orientate and inform visitors on a gallery by gallery basis as well as direct them to individual objects (and in turn provide data such as visitor flow and dwell time back to the museums).

Currently, museums and cultural organisations are just beginning to get an idea of how they might use BLE beacons. The Rubens House in Antwerp, for example, has developed a museum trail, using the beacons to push notification when the visitor is in front of the right object.[5] Visit Bristol and Bristol Museums are involved in the *Shufdy* project, a large-scale trial of 200 beacons located across the city.[6] This includes a treasure hunt element with five hidden beacons that trigger a free copy of a digital magazine.

## 3. Initial Testing

The relative lack of previous work conducted in the area of BLE proved to be a double-edged sword. On the one hand it opened up the opportunity to design and build a truly unique and innovative experience. On the other hand this also meant that we had very little previous work in terms of evaluation data and guidelines to build on. An initial stage of our project thus had to be devoted to researching and evaluating the capabilities offered to us by BLE to better understand its strengths as well as limitations and ultimately translate them into a working solution. A series of experiments was therefore conducted to measure the signal distribution generated by BLE beacons in a museum space. The received signal strength (henceforth RSS) from different beacons naturally varied in different locations of a museum interior. By developing a signal coverage map, our hope was that specific combinations of RSS values could be locked to important locations in the museum space and subsequently used to trigger appropriate content in our location-based game.

Not entirely unexpectedly did our study uncover a steady RSS deterioration whenever moving away from a BLE beacon. This falloff was however found to be rather variable, with irregular fluctuations over time. The nature of the museum interior with large shelves and artefacts scattered throughout the space, proved to be an important factor contributing to this somewhat unpredictable signal coverage. Even the direction facing by the mobile user was frequently found to have a dramatic impact on the RSS. For instance being turned with the back towards the beacon, thus effectively obstructing the path of the signal, could make the RSS drop quite considerably. A seemingly banal factor such as museum visitors randomly traversing the museum space, were found to be another factor reducing the quality of received signal. In fact, larger crowds were even able to block out the signal entirely.

It is however reasonable to assume that all of these issues are an inseparable component of most public environments and as such must be factored in during the design process of any BLE-centred system intended to be operational in this context. Nonetheless the discovered signal strength inconsistencies across time and space posed a significant obstacle to our initial plan of a location-based game in which the position of the user could be accurately estimated by signal strength values from different BLE beacons in the museum space. Due to existing technological limitations of BLE, there is simply no good way of eliminating the risk that the user will be provided with incorrect spatial data due to signal fluctuations. Results of this initial testing phase thus left us with the seemingly intimidating challenge of coming up with a concept that would be operational in spite of these difficulties.

**4. Concept Design**

Many mobile location technologies assume a grid or map metaphor, in which the primary goal is to identify the position of the mobile device and thus the user) within a virtual coordinate system. The most familiar example of this metaphor is provided by the ubiquitous GPS (global positioning system) receivers that are included in most smartphones. In a typical GPS application, the user sees a map or satellite image, with his or her position indicated by a 'pin' at a particular coordinate position on the map. This grid metaphor is reinforced by location service APIs (application programming interfaces) that have access to databases of wireless network locations. Even where the GPS receiver in a device is turned off, or where satellite signals cannot be obtained because the user is indoors, it is now commonplace to see one's location marked as a pin on a map even where the grid coordinates of the device are unknown.

When designing advanced location-based games, there is a temptation to use readily available grid coordinates as the basis for some kind of 'augmented reality', in which the system maintains a coordinate-based virtual world model that must be registered against the real world. The virtual world is taken to be definitive (because it defines the interactive game elements), while the physical world is more-or-less perfectly registered, with reductions in quality of the game experience when the registration error between the physical and virtual worlds becomes too large. One response to this situation is to create new indoor location systems that can determine a fine-grained grid within a building, for example by using ultrasonic transducers and triangulating position relative to their locations.[7] Where this kind of special infrastructure is not available, cruder triangulation can be attempted by techniques such as comparing the signal strength of WiFi hubs or other radio-frequency signal sources. However, these techniques are generally inaccurate. Signal paths can be unpredictable due to reflections and material variations, and signal strength varies considerably in response to the angle at which the device is held, whether it is shielded by the hand, whether other people are passing nearby and so on. As a result, indoor grid positioning systems are either expensive or inaccurate.

There is an alternative to this grid-based metaphor, however. As noted by Weiser,[8] the physical world need not be treated as if it were an extension or imperfect imitation of a virtual model. In contrast, Chalmers[9] proposed that users should be made explicitly aware of gaps in WiFi coverage, or places in which GPS signals cannot be received. Emphasizing such 'seams' draws attention to the physicality of the space that the user occupies, rather than attempting to minimise perceived discrepancies with regard to a virtual environment grid.

This strategy, of drawing attention to physical context rather than digital metadata, also offers a direct application to the ways in which a museum acts as a space for physical engagement. Although museum collections do have digital counterparts, including catalogues, websites, visitor guides and so on, the primary purpose of the museum is the objects themselves, and the space in which they are encountered. By choosing a design strategy that emphasises 'seamfulness' of location-sensing technologies in the manner introduced by Chalmers[9], i.e. by making users aware of all the elements in a physical environment interfering with a smooth functioning of the technology, we are able to emphasise and accentuate this essential physicality of the museum.

With these findings in mind we designed an experimental mobile game named *Ghosts!*. When walking through the museum space with Ghosts! turned on, users would randomly encounter 'ghosts' popping up on the screen of the mobile device explaining to the player that they are lost and need help finding their way back to their home artefacts. Every such home artefact would be equipped with its own BLE beacon. While moving through the museum space, the ghost would then provide users with positive or negative feedback telling them if they're going in the right or wrong direction. The nature of this feedback depends on the RSS received from the BLE beacon of the home artefact. If the RSS value is strong, the ghost becomes happy and provides the user with encouraging messages, such as 'Yes, I can see we're going into the right direction!' On the other hand if the user moves through an area with weak signal coverage, the ghost turns angry and reports that you are getting lost. This means that the deteriorating signal strength over distance was used to estimate how far away from the artefact the user was.

When designing this concept, we also took into consideration various seams, such as fluctuations in the signal coverage. For instance, when the user walks into a large crowd, blocking out much of the BLE signal, the angry ghost response seems appropriate. If the user reacts by raising the smartphone above the crowd, or simply walks into a more open space, the ghost announcing that it can now *see where you're going* again appears as a natural and correct feedback. This means that by taking into consideration potential problems of the BLE technology and by designing them into the actual game, we were thus still able to develop an experience that was functioning reasonably well despite the dynamic nature of a public museum environment.

Moreover, by carefully selecting the artefacts each ghost wants to find and by making the ghosts contact users in a predetermined order, we could push the users towards visiting key artefacts in the exhibition in an appropriate order. This could in turn be used to mediate a narrative or any other experience that would require users to visit artefacts in a specific order.

Once all of these ghost directed tasks are completed, the user receives an achievement for completing the museum and the option to share it through social media, such as Facebook. Finally, users are contacted by a last lost ghost, which this time explains that it comes from a different museum and needs help finding its way back home. This final ghost serves as an invitation to visit other museum spaces and makes users aware that every museum is a brick of a larger experience.

Our prototype application is thus essentially a museum guide masking itself as a hide and seek game. As Falk and Dierking[10] observe, making things happen within the physical environment of the museum provides an important context for the visitor. Instead of layering the digital as an alternative or supplement to the physical gallery space we aim to integrate it into that space. Using the strength, or lack of strength, of the beacon signal to mediate the information presented to the visitor, we hope to involve them in a playful way with moving around and between the museums.

From a museum perspective, what makes this use of Beacon BLE technology particularly intriguing is the ability to create an ongoing dialogue between objects (or their digital representation in a mobile app) and visitors where the communication between the two is constantly changing according to how visitors move through, and around, gallery spaces. So, rather than a direct user action through specific commands or instruction, the game environment we are testing considers user response in the form of movement behaviour as if in conversation with the output from the beacons via object caricatures in the mobile app. Furthermore, in embracing dead spaces and the current geo-location limitations of the BLE technology as a feature of the technical characteristics of the game environment, this allows museums to examine quite closely the ability for beacons to influence visitor movement across large gallery spaces (and potentially different floors) without the need for expansive technical infrastructure.

## 5. Conclusion

Contemporary mobile technologies are expected to operate across an incredibly diverse set of environments. As a result it is inevitable that various 'seams' come in the way of a smooth user experience. As designers, we are thus left with two options. We can be negative about such seams in the system by trying to *design* them out. This might however turn out to be a time and resource consuming process. Our second option is to instead embrace seams such as signal strength inconsistencies. We attempted to argue in favour of the latter option and by turning seams into a driving element of the experience, we attempted to demonstrate the viability of this alternative design paradigm. In the light of our findings it seems reasonable to believe that in the age of ubiquitous computing we cannot hope to achieve desirable results by applying the same design approaches used for traditional digital experiences. Although developed for specific museum environments, we believe that our design approach relying on seamfulness can be applicable to a broad spectrum of uses in the emerging internet of things media landscape, beyond just the museum setting.

____________________________________________________________
## Notes

[1] Mike Kuniavsky, *Smart Things: Ubiquitous Computing User Experience Design* (Morgan Kaufmann, 2010).

[2] Sandeep Kamath and Joakim Lindh, *Measuring Bluetooth Low Energy Power Consumption* (Application Note AN902, 2012), 1—23.

[3] University of Cambridge, 'Become a Card-Carrying Fan of Cambridge University's Eclectic Collections', *cam.ac.uk*, April 30, 2009, viewed on 1 August 2014, http://www.cam.ac.uk/news/become-a-card-carrying-fan-of-cambridge-university%E2%80%99s-eclectic-collections. Blog.

[4] University of Cambridge, 'Fitzwilliam Trials New Visitor Guides', *cam.ac.uk*, July 24, 2002, viewed on 1 August 2014, http://www.cam.ac.uk/news/fitzwilliam-trials-new-visitor-guides  Blog.

[5] Prophets, 'iBeacon Brings Museum to Life', *Prophets.be*, February 2, 2014, viewed on 1 August 2014, http://www.prophets.be/ibeacon-brings-museum-to-life/. Blog.

[6] CPA Group, 'Shufdy', *cpagroup.co.uk*, 2014, viewed on 1 August 2014, http://cpagroup.co.uk/shufdy  Blog.

[7] Mike Hazas and Andy Ward, 'A Novel Broadband Ultrasonic Location System,' in *Proceedings of UbiComp 2002 Fourth International Conference on Ubiquitous Computing* (2002), 264–280.

[8] Mark Weiser, *The World Is Not a Desktop* (Interactions, 1994).

[9] Matthew Chalmers, 'Seamful Design and Ubicomp Infrastructure,' in *Proceedings of Ubicomp 2003: At the Crossroads: The Interaction of HCI and Systems Issues in UbiComp* (2003),

[9] Matthew Chalmers, 'Seamful Design and Ubicomp Infrastructure,' in *Proceedings of Ubicomp 2003: At the Crossroads: The Interaction of HCI and Systems Issues in UbiComp* (2003),

[10] John Falk and Lynn Dierking, *The Museum Experience* (Washington D.C.: Whalesback Books, 1992).

**Tommy Nilsson** has recently graduated with a master's degree in interaction design at the Umeå University. His degree thesis, which he wrote at the University of Cambridge under the supervision of Professor Alan Blackwell, described a seamful approach to design of mobile-based BLE applications and laid ground for the project described in this chapter.

**Alan Blackwell** is Reader in Interdisciplinary Design at the Cambridge Computer Laboratory, having prior qualifications in professional engineering, computing and experimental psychology. He is a fellow of Darwin College, and co-founder and director of the Crucible network for research in interdisciplinary design.

**Carl Hogsden** is a Digital Projects Research Associate working for the University of Cambridge Museums. As a museum practitioner, technologist and developer his prime interest is in ways that online collaboration can enable a conduit between the internet and physical dimensions of the museum for integrating multi-vocality in collections-based activity**.**

**David Scruton** is the Documentation and Access Manager at the Fitzwilliam Museum, with a particular interest in the use of collections information and the impact of new technology on the relationship between museums and their audiences. With a research background in art history he has worked previously in organising exhibitions of contemporary art and new media.